# Hardware-Friendly Synaptic Orders and Timescales in Liquid State Machines for Speech Classification

Vivek Saraswat, Ajinkya Gorad, Anand Naik, Aakash Patil and Udayan Ganguly

*Dept. of Electrical Engineering*
*Indian Institute of Technology, Bombay*, Mumbai, India
udayan@ee.iitb.ac.in

*Abstract*—Liquid State Machines are brain inspired spiking neural networks (SNNs) with random reservoir connectivity and bio-mimetic neuronal and synaptic models. Reservoir computing networks are proposed as an alternative to deep neural networks to solve temporal classification problems. Previous studies suggest $2^{nd}$ order (double exponential) synaptic waveform to be crucial for achieving high accuracy for TI-46 spoken digits recognition. The proposal of long-time range (ms) bio-mimetic synaptic waveforms is a challenge to compact and power efficient neuromorphic hardware. In this work, we analyze the role of synaptic orders namely: $\delta$ (high output for single time step), $0^{th}$ (rectangular with a finite pulse width), $1^{st}$ (exponential fall) and $2^{nd}$ order (exponential rise and fall) and synaptic timescales on the reservoir output response and on the TI-46 spoken digits classification accuracy under a more comprehensive parameter sweep. We find the optimal operating point to be correlated to an optimal range of spiking activity in the reservoir. Further, the proposed $0^{th}$ order synapses perform at par with the biologically plausible $2^{nd}$ order synapses. This is substantial relaxation for circuit designers as synapses are the most abundant components in an in-memory implementation for SNNs. The circuit benefits for both analog and mixed-signal realizations of $0^{th}$ order synapse are highlighted demonstrating 2-3 orders of savings in area and power consumptions by eliminating Op-Amps and Digital to Analog Converter circuits. This has major implications on a complete neural network implementation with focus on peripheral limitations and algorithmic simplifications to overcome them.

*Keywords—LSM, reservoir, speech classification, SNNs, synapse, order, timescale*

## I. INTRODUCTION

Spiking Neural Networks (SNNs) are the third generation of artificial neural networks [1]. Information is encoded in the timing of spikes of the neurons which results in temporally-specific and low-power communication events in the neural network [2]. SNNs have been adopted for a large number of classification and pattern recognition tasks with a focus on developing dedicated neuromorphic hardware to harness the power efficiency of SNNs [3], [4]. Although, biologically more plausible, SNNs face significant challenges in mapping learning algorithms to spiking neurons [5], [6] as well as in circuit design [7]. Liquid State Machines (LSMs) are an attempt to simplify the network development and learning strategies by borrowing further inspiration from the human cortex. LSM is a spiking neural network architecture which has a reservoir of recurrently connected neurons [8]. This architecture is proposed as an alternative to Deep Neural Networks which comprise of a large number of successive layers of neurons. These layers can differ in size (number of neurons per layer) and the objective is to have sufficient number of learning parameters (inter-layer weights and neuronal biases) so as to achieve arbitrary classification functionality. LSMs, on the other hand, deviate from this depth of network idea. In the reservoir computing framework, the presence of recurrent dynamics and higher dimensionality of information represented in the reservoir is entrusted with facilitating arbitrary separation and generalization functions necessary for classification tasks. LSMs have been shown to be particularly well suited for temporal information datasets like speech and video activity recognition [9]–[12]. The reservoir response is expected to act as a universal function generator and all learning is pushed to a single linear classifier outside the reservoir itself [8].

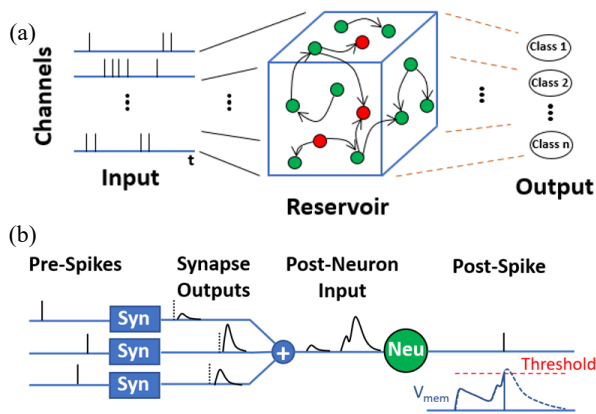

Fig. 1. (a) LSM Architecture – spiking input channels, randomly connected reservoir of excitatory (green) and inhibitory (red) neurons and output classifier neurons, the input and reservoir weights are fixed while the classifier weights are trained, (b) Basic computations in a reservoir of leaky-integrate and fire (LIF) neurons: Pre-spikes are scaled and shaped by synapses, then summed and integrated by the post-neuron, a spike is issued when membrane potential reaches the threshold and the potential is reset

Figure 1 shows an example LSM. There is an input layer of spiking neurons (called input channels), a large reservoir of recurrently connected neurons and an output layer (Fig. 1(a)). The input spikes are generated from the raw input features using a preprocessing step which is specific to the dataset. The reservoir connectivity follows cortical brain-inspired random

This work has been supported by Prime Minister's Research Fellowship, Ministry of Electronics and Information Technology, Govt. of India. Ajinkya Gorad is currently affiliated with Aalto University, Finland.

connectivity. Different models exist like the local probabilistic connections, axonal model and small-world networks [13]. Once generated using a random model, the reservoir connections are not changed during the network operation. A single linear classifier between the reservoir neurons and the output neurons is chosen to be trained. The basic computations that happen in a recurrent network are shown in Fig. 1(b). The spikes issued by different pre-neurons arrive at synapses of these neurons with the post neuron. Each synapse outputs a post-synaptic current in response to the pre-neuron spike after a small fixed delay. The synapses output a scaled current waveform. The amplitude depends on the connection strength of the synapse. The summed current is then integrated by the post-neuron which is typically modelled as a leaky integrate and fire neuron. An output spike is issued by the post neuron if the integrated membrane potential reaches a threshold and the potential is reset to resting value.

A recurrent neural network (like the reservoir) is best implemented as an in-memory architecture where neurons are able to interact with each other by means of the more numerous synaptic connections in a parallel manner [14]–[16]. The model proposed for spiking neurons is typically the leaky-integrate and fire neurons [17]. There are different models of the post-synaptic waveforms of the unit strength synapse in response to an input spike (Fig. 2): (a) $\delta$ synapse outputs high current for a single time step, (b) 1st order synapse outputs an exponential decaying current and (c) 2nd order synapse outputs double exponential rise and fall waveform. Typically, the biomimetic 2nd order synapse waveform is used by the biologically inspired algorithms [18]. Recently, a digital implementation of an LSM using these models was proposed for the TI-46 spoken digits recognition task with a spike based local learning rule for the linear classifier [17]. Further, the LSM network was represented using a state-space model and a performance predicting memory metric was extracted [11]. It has been argued that the post synaptic current waveform has a crucial role to play in the classification accuracy [18]–[20] specifically for the speech digit recognition task [17] (Table I). Error was shown to degrade by more than 10 times if a $\delta$ synapse is employed in place of the 2nd order synapse waveform keeping the connection strengths unchanged [17]. This immediately renders the circuit-friendly $\delta$ synapse unfeasible for circuit designers implementing SNNs. Complex waveforms of higher (1st and 2nd) order synapses are not circuit-friendly [21]. The area and power consumed by circuits realizing long-time (ms) range waveforms especially relevant for real-time sensory input data is a challenge for neuromorphic chip designers [22].

Algorithmic simplifications to waveforms requirement can have significant impact for hardware realizations of SNNs. Hence, in this paper, first we show the strategy to obtain the optimal operating point and its relation with reservoir spiking activity for a given unit synapse waveform and timescale. For this we perform a more comprehensive parameter sweep on the connection strengths for the TI-46 spoken digits recognition task. Next, we propose a 0th order synapse (Fig. 2(d)) with a timescale which is essentially a rectangular pulse of finite width to get the best of both performance and circuit friendly implementation. $\delta$ synapse is a special case of the 0th order synapse where pulse width equals a single time step. Finally, we perform a circuit cost and benefits estimation for the different synaptic orders that greatly affects the circuit design choices by the SNN chip designers.

TABLE I. PERFORMANCE OF DIFFERENT SYNAPSE ORDERS [17]

| Metric | $\delta$ Synapse | 1st Order Synapse | 2nd Order Synapse |
|---|---|---|---|
| Accuracy (%) | 88.85 | 90.73 | 99.09 |
| Error (%) | 11.15 | 9.27 | 0.91 |
| Feature | Circuit-friendly | Biomimetic | Biomimetic |

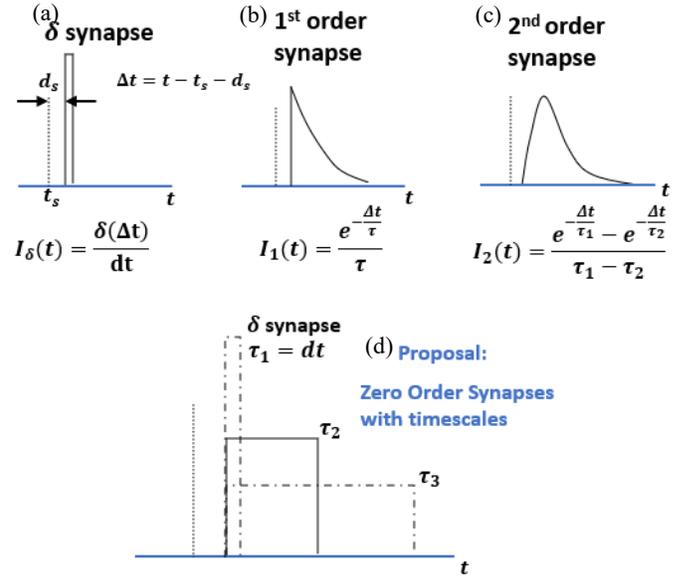

Fig. 2. Different types of unit synapses: (a) $\delta$ (b) 1st order (c) 2nd order synapses. Dotted line is the spike time, $t_s$, in response to which the synaptic current waveform is output after a delay, $d_s$. (d) Proposed 0th order rectangular synapse with finite pulse width. A synapse can have different timescales ($\tau_1, \tau_2, \tau_3$) but the total integrated current is assumed constant for a unit synapse. A $\delta$ synapse is a special case of 0th order synapse where pulse width is equal to a single time-step dt.

## II. TI-46 SPOKEN DIGITS RECOGNITION SETUP

The TI-46 spoken digits dataset comprises of 5 speakers uttering 10 times each of the 10 digits (500 samples) [17]. Each utterance is about 1 second long. The preprocessing of the audio waveforms to achieve a spiking input for the LSM is well established and follows the Lyon's Passive Ear Model [23]. This is a human ear Cochlea inspired model which works with 77 channels of band-pass filters followed by Automatic Gain Control and resampling stages. Finally, Bens Spiking Algorithm employs a rate coding-based scheme to output spike trains in 77 channels for each input digit sample. The details of the preprocessing stage have been discussed in detail elsewhere [11], [24]. These 500 samples are used for training and testing the network in a 5-fold manner. 400 samples are used to train the network and remaining 100 are tested in a rotating manner (Fig. 3).

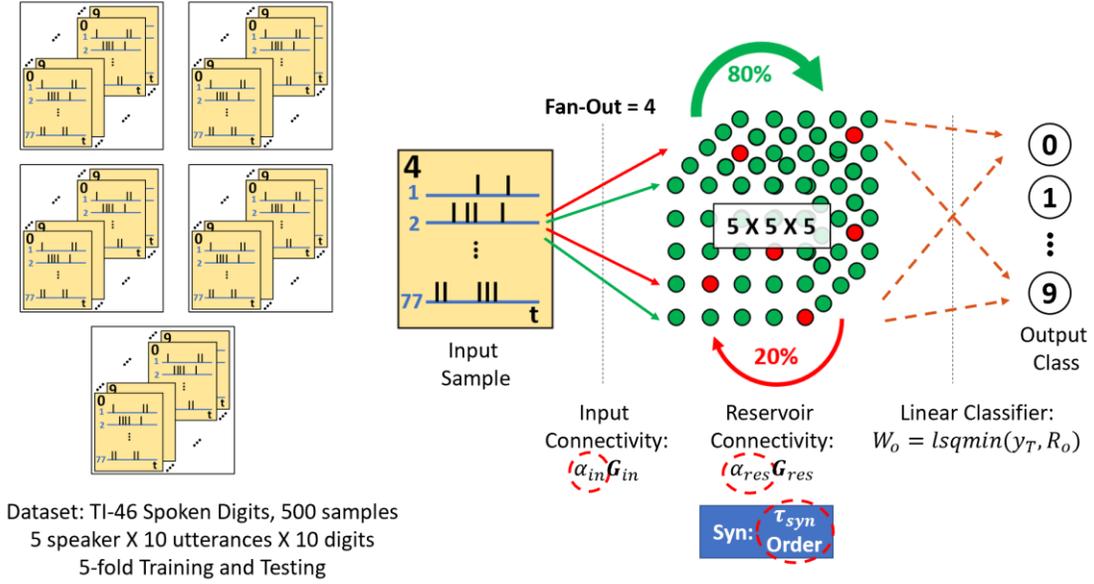

Fig. 3. TI-46 preprocessed dataset and spoken digit recognition setup. The quantities encircled by red dashed lines are the design space parameters for this study.

The reservoir consists of 125 neurons. The reservoir connectivity ($G_{res}$) is established using the local probabilistic random connections model with 80% excitatory neurons and 20% inhibitory neurons [11]. This model requires the neurons to be arranged on a grid (5 X 5 X 5 chosen here) to calculate the connection probability as a function of the distance between the neurons in the grid structure. The exact parameters for reservoir generation are discussed in detail in [11]. Each spike train channel of any input digit sample is connected to 4 randomly chosen reservoir neurons with equal number of excitatory and inhibitory connections of equal strength. This results in an input connectivity matrix ($G_{in}$) (Fig. 2). Although, the connectivity matrices $G_{in}$ and $G_{res}$ are not changed, a constant scaling factor to synaptic strengths $\alpha_{in}$ and $\alpha_{res}$ are introduced as tuning parameters. These decide the operating point of the network for any given synapse model. The reservoir output spikes feed into the output neurons using the classifier weights. The reservoir output spikes for all neurons are averaged over the entire duration of the sample to get the reservoir output response for training. This reservoir response for all training samples and the target output layer response is used to train the classifier weights by a least squares minimization algorithm:

$$W_o = \min_{W}||Ky_T - WR_o||^2 \quad (1)$$

where $W_o$ are the optimal output weights (artificially limited to $\pm W_{lim}$), $K$ is a scaling constant, $y_T$ is the expected target spiking in the output layer collated for all training samples and $R_o$ is the time-averaged reservoir output response. The trained weights are then tested on the testing samples to calculate the speech classification accuracy by identifying the maximum spiking activity output neuron.

The neurons are modelled as identical Leaky Integrate and Fire neurons. The synaptic order and timescale are also identical for all synapses in the network. However, we have the option of choosing the synapse order and timescale of the unit synapse apart from the previously mentioned input and reservoir weights scaling, $\alpha_{in}$ and $\alpha_{res}$, in this simulation setup (Table II).

TABLE II. SIMULATION SETUP

| Component | Parameter | Values |
|---|---|---|
| Preprocessing [Zhang et al.] [17] | Model | Lyon's Passive Cochlea |
| Reservoir [Gorad et al.] [11] | Model | Local Probabilistic (80E/20I) |
| | $\alpha_{in}$ | 1 – 20 |
| | $\alpha_{res}$ | 0.1 – 4 |
| Synapse | Order | $\delta$, 0, 1, 2 |
| | Timescale $\tau$ | 1 – 50 ms |
| Neuron | Leakage $\tau_N$ | 64 ms |
| | Refractory Period | 2 ms |
| | Threshold | 20 mV |
| | Resting | 0 mV |
| Classifier | Model | Least Squares |
| | $W_{lim}$ | 8 |
| | $K$ | 1000 |

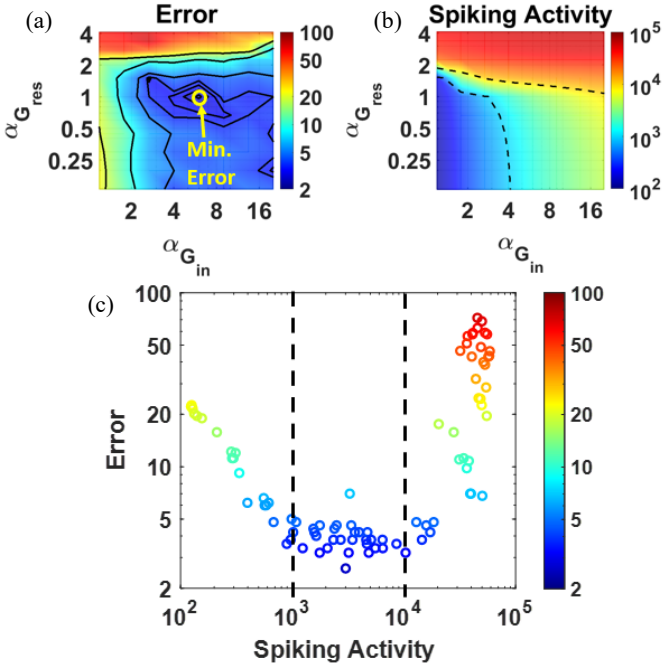

Fig. 4. Experiments for 2nd order synapse with timescales $(\tau_1, \tau_2) = (8\text{ ms}, 4\text{ ms})$ – the fall and rise timescales. (a) Error as a function of the input and reservoir weights scaling – minimum error is highlighted and equi-error contours are plotted, (b) Spiking activity in the reservoir averaged over all the input samples as a function of the input and reservoir weights scaling, (c) Error as a function of the spiking activity in the reservoir shows an optimal range of reservoir spiking activity (denoted by dashed lines)

## III. RESULTS AND DISCUSSION

### A. Effect of weights scaling

In order to evaluate the peak performance of a given synaptic order and timescale, we perform a parameter space sweep for $\alpha_{in}$ and $\alpha_{res}$. There is an optimal $\alpha_{in}$ and $\alpha_{res}$ point where the 5-fold average classification accuracy is maximum or the error is minimum (Fig. 4(a)). The weight scaling parameters jointly control the level of spiking activity in the reservoir, i.e., spikes issued by all reservoir neurons averaged over all the input samples. As expected, the reservoir spiking activity rises monotonically with $\alpha_{in}$ and $\alpha_{res}$ (Fig. 4(b)). The response w.r.t $\alpha_{res}$ is more non-linear due to the recurrent nature of the reservoir connections setting up a positive feedback and chaotic dynamics [11]. A more insightful representation is to observe the variation of the error with the spiking activity of the reservoir (which takes into account the net effect of both the weights scaling). Error is minimum for an optimal range of spiking activity (Fig. 4(c)). The error rises rapidly above and below this optimal spiking activity range. Thus, the peak performance for a given synaptic order and timescale is evaluated from its error vs spiking activity graph. This is aligned with the well-known "edge of chaos" theory for maximum performance [11]. The observed optimal spiking activity range depends on the actual application and the associated network topology. Nonetheless, a non-monotonic trend in the achieved error rates is expected with the spiking activity in general. Hence, this framework provides a method to bias the reservoir in a high-performing region. Although, the optimal operating point has been achieved by tuning the weights scaling, ultimately, the effect of this tuning is a modulation of the reservoir spiking activity. In practice, this may be achieved not just by synaptic plasticity but also by altering neuronal excitability and/or timescales. Efficient hardware implementations of controllable synaptic input integration neurons [25], [26] and learning synapses [27], [28] is key and can benefit LSMs hugely.

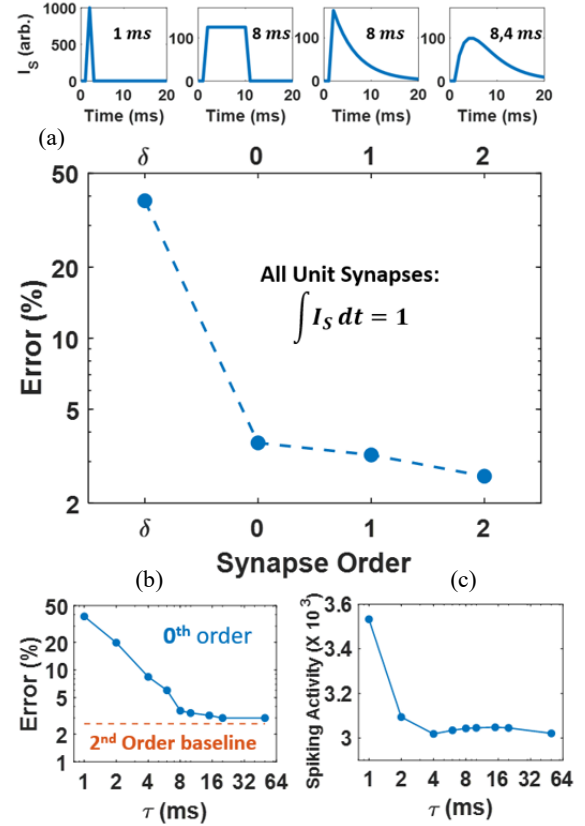

Fig. 5. Fixed $(\alpha_{in}, \alpha_{res}) = (6, 1)$ simulations for (a) Error as a function of the synapse order; synaptic timescales of different orders are as shown in the inset at the top (synapses follows constant integrated current model when order or timescale is varied), (b) Error as a function of the synaptic timescale of the 0th order synapse and comparison with the 2nd order $(\tau_1, \tau_2) = (8\text{ ms}, 4\text{ms})$ baseline, (c) Spiking activity in the reservoir as a function of the synaptic timescale of the 0th order synapse

### B. Effect of synapse order and timescale

The optimal $\alpha_{in}$ and $\alpha_{res}$ determined from the 2nd order synapse experiments are now used for testing the other unit synaptic orders. Whereas the $\delta$ synapse shows a marked deterioration in error, the 0th and the 1st order synapses are still comparable to 2nd order performance provided they have comparable timescales (Fig. 5(a)). This is significant since 0th order synapses are much more circuit-friendly than higher order synapses as discussed in Section IV. Next, we perform a synapse timescale variation study for the 0th order synapse for the fixed $\alpha_{in}$ and $\alpha_{res}$ (Fig. 5(b)). $\delta$ synapse is a special case of the 0th order synapse where pulse width equals a single time step (1 ms). When the synaptic timescale is varied, the integrated current is kept unchanged (constant charge or unit synapse variation). This leads to smaller timescales affecting the membrane potentials of the leaky integrating neurons more significantly and a sharp rise in the spiking activity of the reservoir for the $\delta$ synapse (Fig. 5(c)). As identified earlier this increased spiking activity is accompanied by a reduced classification accuracy for the $\delta$ synapse. However, when the synaptic timescale of the $0^{th}$ order synapse is comparable to the optimal $2^{nd}$ order synapse timescale (~ 8 ms), the accuracies achieved by both the synapses are very similar (Fig. 5(b)). Hence, the $0^{th}$ order synapse is as high performing as the second order synapse for the inference tasks if the timescale is allowed to be tuned.

*C. Peak Performance of all synaptic orders and timescales*

We observe that the previous prediction of $\delta$ synapse with fixed $\alpha_{in}$ and $\alpha_{res}$ being unsuitable for the speech classification LSM is a limited experiment (Table I and Fig. 5(a)). Hence, we next calculate the peak performance of different synaptic orders and timescales according to the strategy highlighted in Section III.*A*. We allow a scan over the $\alpha_{in}$ and $\alpha_{res}$ parameters to identify the best average 5-fold accuracy for a given unit synaptic waveform and then repeat this for $0^{th}$, $1^{st}$ and $2^{nd}$ order synapses of different timescales (Fig. 6(a)). For $2^{nd}$ order, the $\tau$ plotted is the fall $\tau$ with the rise $\tau$ being one-half of the fall $\tau$. A more comprehensive method of obtaining optimal operating point demonstrates that even $\delta$ synapses are not much worse off compared to the optimal timescale $2^{nd}$ order synapse (Table III). The peak performances for all orders and timescales are correlated to an optimal range of reservoir spiking activity (Fig. 6(b)). Thus, the optimal operating point strategy shows the true performance utility of different synapse models in contrast to previous studies (Fig. 7). This opens an avenue for the circuit designers to opt for $0^{th}$ order synapses as a realistic choice for hardware implementation of such biologically inspired algorithms. The $0^{th}$ order synapse have numerous circuit implementations related benefits which we discuss next.

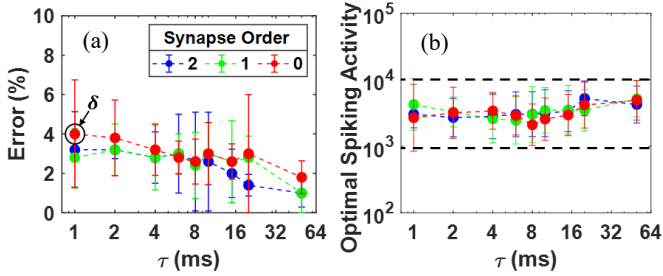

Fig. 6. (a) Minimum Error as a function of the synapse timescale for different synaptic orders calculated by sweeping weights scaling parameters to obtain the optimal operating point, error bars are plotted using the standard deviation of the 5-fold testing data per simulation. (b) Optimal reservoir spiking activity (mean and error bars) for the 10 lowest Error operating points during weights scaling sweep as a function of the synapse timescale for different synaptic orders validating the optimal spiking range framework.

TABLE III. COMPARISON OF TI-46 SPOKEN DIGIT CLASSIFICATION

| Work | Synapse Order | Operating Point | Accuracy (%) |
|---|---|---|---|
| Verstraeten et al. [24] | Biomimetic $2^{nd}$ order | - | 98 |
| Zhang et al. [17] | Circuit-friendly $\delta$ | Does not change with synapse order and $\tau$ | 89 |
| | Biomimetic $1^{st}$ order | | 91 |
| | Biomimetic $2^{nd}$ order | | 99 |
| This Work | Circuit-friendly $\delta$ | Optimal reservoir spiking activity strategy | 96 |
| | Circuit-friendly $0^{th}$ order | | 98 |
| | Biomimetic $1^{st}$ order | | 99 |
| | Biomimetic $2^{nd}$ order | | 99 |

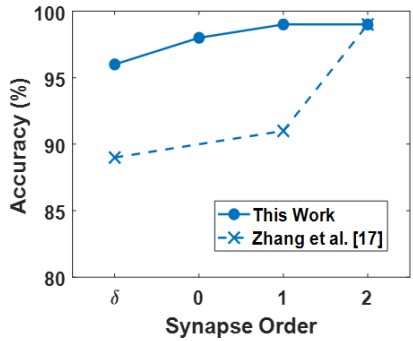

Fig. 7. Optimal classification accuracy as a function of the synapse order for the reservoir spiking activity based optimal weights scaling operating point proposed in this work compared to the fixed operating point results of previous works

## IV. CIRCUIT COST ESTIMATION

Spiking Neural Networks are a fundamentally different computing architecture with integrated memory and computational units and temporal information encoding. Whereas the conventional von-Neumann microprocessors and memories scale in terms of speed and area, they are primarily serial processors with each of the components (ALUs, cache) speeding up in time to improve overall throughput and efficiency.

Neuromorphic chips, which implement SNNs, tend to focus on radically different aspects. The computation is massively parallel and often an in-memory approach is used to accelerate the temporal vector-matrix multiplication [14]–[16]. The neuronal outputs are scaled and summed to generate the next stage inputs in a parallel manner. The weight matrix or the connection matrix becomes the most dominant circuit. An efficient connection unit or the synapse, hence, plays a critical role in determining how much power and area consuming, the SNN chip is going to be. This becomes a circuit design issue compounded by the fact that many real-time temporal classification tasks like speech processing require the implementation of long timescales (in the range of milliseconds) for the neuron's leakage and synaptic waveforms. Longer timescales are linked to increased design capacitance and hence lead to higher circuit area and power consumption [22]. This issue is typically overcome by operating the CMOS circuits in subthreshold regimes or making use of novel physics like impact ionization or band-to-band tunneling based neurons and synapses [25]–[27], [29], [30]. Both these approaches are under extensive investigation and show immense promise however they are plagued with variability and latency issues [29], [31]. It is possible that the SNN based algorithms may be robust to such component level variabilities if they employ a feedback mechanism [32]. Nonetheless, circuit-friendly simplifications to biologically motivated algorithms can go a long way in allowing conventional CMOS circuits to be well suited for designing neuromorphic chips.

One such simplification is proposed in this paper: the synapse order. Previously proposed algorithms for spoken digits classification have claimed a 2nd order synapse to be crucial for performance [17]. However, as shown in Section III.C, for any choice of synapse order and timescale, the optimal operating point requires a design space exploration on weights scaling to achieve optimal reservoir spiking activity. As a result, the degradation in performance by choosing a 0th order synapse is demonstrated to be much lower compared to previous claims (Table III and Fig. 7). Furthermore, the absolute accuracy for the $\delta$ synapse is also high (96 %). The accuracy further rises (~ 98 %) for the 0th order synapses if the timescale of the rectangular pulse is allowed to increase and become comparable to 2nd order pulse timescales (Fig. 6(a)). This knowledge addition drastically impacts the choices of circuit designers now. The 0th order synapses are feasible and provide excellent performance. Further, the manifold circuit benefits that accompany circuit implementation of the 0th order synapse compared to 2nd order are presented next:

### A. Benefits for analog implementation

Analog waveform shaping involves charging/discharging large capacitors followed by signal buffering circuits as drivers [22], [27], [33]. For e.g. to generate timescales in the range of milliseconds using minimum sized 45 nm CMOS technology transistors biased in the subthreshold regime requires a capacitance of 150 fF which is about 100 times the gate capacitance of the minimum sized transistor in that technology [34], [35]. This indicates that the majority of the synapse area will be occupied by the waveform shaping circuitry primarily the capacitance. As the number of timescales to be implemented in the waveform increase (as in higher order synapses), so do the number of capacitances. It is possible to achieve second order waveforms using single capacitances by differential pair integrator (DPI) circuits [19], [36] (Table IV). However, they have increased circuit complexity and bias generation circuits requirement rendering them vulnerable to device variability. Also, any analog signal needs to be buffered through a driver circuit to drive the next stages in the network without loading itself. A general analog waveform can be buffered using an operational amplifier which is a big (~ 1000 $\mu m^2$ in 0.25 um CMOS technology node) and complex circuit affected by static bias power consumption (~ 100-500 $\mu W$) and variability [37], [38] (Table V). A binary level digital signal like a 0th order synapse, on the other hand, can be buffered much easily by a small (~ 10 $\mu m^2$ in 0.25 $\mu m$ node) inverter pair in series eliminating the need of a more complex drive circuitry. Since digital buffers only consume dynamic power with minimal static leakage, the power consumption (~ $C_{load}V_{DD}^2/\tau$) at 1 ms timescale range is greatly reduced (~ 100 nW for 4 pF $C_{load}$ and 3.3 V supply) compared to the analog buffer [39].

In addition, any rise and fall time of an analog waveform is determined by both the charging/discharging timescale and the location of the steady-state signal value w.r.t the present signal value. This means that constraints on the rise and fall times (as in higher order synapses) require the steady state voltage or the $V_{DD}$ to be farther away from the unit synapse waveform amplitude. This introduces a voltage margin requirement. Contrary to this, on/off pulses for 0th order synapses without specific constraints on charging/discharging rates can charge all the way to $V_{DD}$ and GND without the requirement of a voltage margin. Thus, analog circuit design for 0th order synapses has significant benefits compared to higher order synapses.

### B. Benefits for mixed-signal implementation

In order to implement a waveform digitally, the waveform needs to be quantized at some levels (say 4-bits for sufficient precision). These levels are stored in a look-up table (LUT) which is a memory bank. In-memory multiply and accumulate operation depend on the synapse outputs to be analog currents that can be summed together in a parallel manner regardless of whether the neurons or synapses themselves are digital or analog implementations. Hence, at the time of application of the synapse output waveform, a digital-to-analog converter (DAC) is required (Table IV). The DAC is a complex mixed signal circuit whose power (500 $\mu W$ for 4-bit and 0.25 $\mu m$ node) and area (~ 1500 $\mu m^2$) scale badly with number of bits as may be the requirement for a high-fidelity reconstruction of the desired waveform [40]–[42] (Table V). A 0th order synapse or a rectangular pulse of finite width, on the other hand, is much easier to implement using binary levels digital counters comprising of simple flip-flops (~ 20 transistors per flip-flop) with minimal circuit complexity (500 $\mu m^2$ for 4-bit i.e., around 80-100 transistors in 0.25 $\mu m$ node) and power consumption (5 $\mu W$ – estimated as 50X (same as size ratio) with the inverter pair dynamic power) [39].

TABLE IV. CIRCUIT COST ESTIMATION

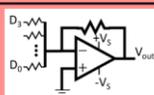

| Category | Synapse Order / Metric | 0th order | Higher order |
|---|---|---|---|
| Performance | Accuracy (%) | 98 | 99 |
| | Error (%) | 2 | 1 |
| Analog Implementation | Timescale Realization | 1C (non-DPI) | 1 or more C (non-DPI) or 1C and 3 biases (DPI) |
| | Driver Circuit | Series inverter pair | Op-Amp |
| | Voltage Margin Required | No | Yes |
| Digital Implementation | Quantization | Binary levels | 4-bit 16 levels LUT |
| | Pulse generation | Counter | DAC |

TABLE V. CRITICAL COMPONENTS: POWER AND AREA COMPARISON

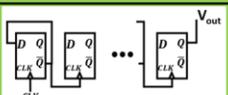

| Component | Higher Order | 0th order | Benefit |
|---|---|---|---|
| Driver [36][38] | Analog Buffer - OpAmp / Power: 100-500 $\mu W$ (Static) / Area: 1000 $\mu m^2$ | Series Inverter Pair / Power: 100 nW (Dynamic) / Area: 10 $\mu m^2$ | 1000 X / 100 X |
| Pulse Generation [39][38] | DAC / Power: 500 $\mu W$ / Area: 1500 $\mu m^2$ | Counter / Power: 5 $\mu W$ / Area: 500 $\mu m^2$ | 100 X / 3 X |

All values are estimated for 0.25 $\mu m$ CMOS technology node 3.3V supply.
Inverters and Analog buffers are assumed to drive 4pF load capacitance.
All dynamic signals are assumed to be in ms (or 1kHz) range.
4bit Counters and 4bit DACs are assumed.

## V. Conclusion

In this paper, we performed a comprehensive parameter space exploration to show the correlation of the peak performance with an optimal reservoir spiking activity range for different synaptic orders and timescales in LSMs for the speech classification task. We demonstrated the true impact on the performance of the $\delta$ synapse (96 %) compared to $2^{nd}$ order synapse (99 %). We proposed the utility of $0^{th}$ order synapses that are algorithmically a feasible choice (98 %) and practically an excellent circuit choice with numerous implementation related benefits. The elimination of large Op-Amps and power hungry digital-to-analog converter (DAC) circuits result in 2-3 *orders* of power and area savings when using $0^{th}$ order synapses. Algorithmic simplifications to biologically plausible networks reduce the circuit burden while retaining the high performance of the classification task.

## Acknowledgment

VS thanks Maryam Shojaei Baghini and Ajay Singh for insightful discussions on circuit choices.